\documentclass[prl,twocolumn,superscriptaddress]{revtex4-1}
\usepackage{}
\usepackage{amssymb}
\usepackage{amsfonts}
\usepackage{graphics,graphicx,epsfig,bm,amsmath,amsthm,amssymb}
\usepackage{bm}
\usepackage{bbm}
\usepackage{longtable}
\usepackage{multirow}
\usepackage{array}
\usepackage{color}
\usepackage[usenames,dvipsnames]{xcolor}

\usepackage{float}
\usepackage[a4paper,colorlinks=true,
linkcolor=blue,citecolor=blue,
pdfauthor={ },
pdftitle={ },
pdfsubject={ },
pdfkeywords={ }]{hyperref}

\begin{document}

\title{Dynamical quantum phase transitions in non-Hermitian lattices}

\author{Longwen Zhou}
\email{zhoulw13@u.nus.edu}
\affiliation{Department of Physics, National University of Singapore, Singapore 117543}

\author{Qing-hai Wang}
\affiliation{Department of Physics, National University of Singapore, Singapore 117543}

\author{Hailong Wang}
\affiliation{Division of Physics and Applied Physics, School of Physical and Mathematical Sciences, Nanyang Technological University,	Singapore 637371}

\author{Jiangbin Gong}
\email{phygj@nus.edu.sg}
\affiliation{Department of Physics, National University of Singapore, Singapore 117543}


\begin{abstract}
In closed quantum systems, a dynamical phase transition is identified
by nonanalytic behaviors of the return probability as a function of
time. In this work, we study the nonunitary dynamics following quenches
across exceptional points in a non-Hermitian lattice realized by optical
resonators. Dynamical quantum phase transitions with topological signatures
are found when an isolated exceptional point is crossed during the
quench. A topological winding number defined by a real, noncyclic
geometric phase is introduced, whose value features quantized jumps
at critical times of these phase transitions and remains constant
elsewhere, mimicking the plateau transitions in quantum Hall effects.
This work provides a simple framework to study dynamical and topological
responses in non-Hermitian systems.
\end{abstract}
\maketitle

\emph{Introduction}.--
Dynamical quantum phase transitions (DQPTs) are characterized by nonanalytic
behavior of physical observables as functions of time~\cite{DPTRev1,DPTRev2}. These transitions
happen in general if the system is ramped through a quantum critical
point. As a promising framework to classify quantum dynamics of
nonequilibrium many-body systems, DQPTs have been studied intensively
in recent years~\cite{PollmannPRE2010,HeylPRL2013-1D,KarraschPRB2013-1D,HeylPRL2014,CanoviPRL2014,VajnaPRB2014,HickeyPRB2014-1D,AndraschkoPRB2014-1D,KrielPRB2014-1D,HeylPRL2015,SharmaPRB2015-1D,HeylPRB2016_1,HeylPRB2016_2,HuangPRL2016,AbelingPRB2016-1D,KarraschPRB2017-1D,LiarXiv2017}.
The generality and topological feature of DQPTs were
demonstrated in both lattice and continuum systems~\cite{FogartyNJP2017-1D}, across different
spatial dimensions~\cite{VajnaPRB2015-1D2D,SchmittPRB2015-2D,WangarXiv2017-2D,UtsoPRB2017-2D,HeylarXiv2017-3D}, and under various dynamical protocols~\cite{DiffRamp1,DiffRamp2,DiffRamp3}. The defining
features of DQPTs have also been observed in recent experiments~\cite{DPTExp1,DPTExp2,DPTExp3}.

Following the initial proposal, most studies on DQPTs focus on closed
quantum systems undergoing unitary time evolution. Efforts have been
made to generalize DQPTs to systems prepared in mixed states~\cite{Mixed1,Mixed2}. However,
DQPTs in systems with gain and loss, and therefore subject to nonunitary
evolution are largely unexplored. One such class of open systems can
be descried by a non-Hermitian Hamiltonian. This type of system, realizable in various platforms like
photonic lattice~\cite{NHPhot1}, phononic media~\cite{NHPhon1}, LRC circuits~\cite{NHCir1} and cold atoms~\cite{NHCoAt1,NHCoAt2}, has attracted great attention in recent years due to their
nontrivial dynamical~\cite{NHDym1,NHDym2,NHDym3,NHDym4,NHDym5,NHDym6,NHDym7,NHDym8,NHDym9,NHDym10}, topological~\cite{NHTop1,NHTop2,NHTop3,NHTop4,NHTop5,NHTop6,NHTop7,NHTop8,NHTop9,NHTop10,NHTop11,NHTop12,NHTop13,NHTop14} and transport properties~\cite{NHTra1,NHTra2,NHTra3,NHTra4,NHTra5,NHTra6,NHTra7,NHTra8}. Many of these features
can be traced back to non-Hermitian degeneracy (\emph{i.e.} exceptional) points
mediating gap closing and reopening transitions on the complex plane~\cite{ExcPot1,ExcPot2,ExcPot3,ExcPot4,ExcPot5,ExcPot6}.
In this work, we explore DQPTs in non-Hermitian
systems, with a focus on topological signatures in nonunitary evolution
following quenches across exceptional points (EPs).

\emph{Theory.}--
We start by summarizing the theoretical framework of DQPTs for systems
described by non-Hermitian lattice Hamiltonians. The nonunitary
time evolution of the system is governed by a Schr\"odinger equation $i\frac{d}{dt}|\Psi(t)\rangle=H|\Psi(t)\rangle$.
For concreteness, we present the formalism with a one-dimensional
two-band lattice model in mind, while the generalization to multiple-band
systems is straightforward~\cite{supp1}. 

Consider a system described by Hamiltonian $H=H^{f}\theta(t)+H^{i}\theta(-t)$,
where $\theta(t)$ is the step function. At time $t=0$, the system
undergoes a sudden quench with its Hamiltonian switched from $H^{i}$
to $H^{f}$. At time $t>0$, the return amplitude of the system to
its initial state $|\Psi^{i}\rangle$ is given by $G(t)=\langle\Psi^{i}|e^{-iH^{f}t}|\Psi^{i}\rangle$.
In our study, the post-quench Hamiltonian $H^{f}$ is non-Hermitian,
and the evolution operator $e^{-iH^{f}t}$ is therefore nonunitary.
Nontrivial dynamics is expected if $|\Psi^{i}\rangle$ is not a right
eigenvector of $H^{f}$. Specially, when the evolving state $|\Psi^{i}(t)\rangle=e^{-iH^{f}t}|\Psi^{i}\rangle$
becomes orthogonal to the initial state at some critical time $t_{c}$,
we have $G(t_{c})=0$. Then the rate function of the return amplitude
$f(t)=\lim_{N\rightarrow\infty}\frac{-1}{N}\ln G(t)$ becomes nonanalytic
at $t=t_{c}$, where $N$ is the number of degrees of freedom of the
system. Similar to DQPTs in unitary evolution, we identify the vanishing
of $G(t)$ at critical times as signatures of DQPTs in non-Hermitian
systems.

In momentum space, the pre-/post-quench Hamiltonian $H^{i/f}$ can
be expressed as $H^{i/f}=\sum_{k}{\cal H}_{k}^{i/f}|k\rangle\langle k|$,
where $k\in[0,2\pi)$ is the quasimomentum and ${\cal H}_{k}^{i/f}=d_{x}^{i/f}(k)\sigma_{x}+d_{y}^{i/f}(k)\sigma_{y}+d_{z}^{i/f}(k)\sigma_{z}$ is the Bloch Hamiltonian,
with $\sigma_{x,y,z}$ being Pauli matrices in their usual representation.
The right eigenvectors $|\psi_{k}^{i/f\pm}\rangle$ of ${\cal H}_{k}^{i/f}$
satisfy ${\cal H}_{k}^{i/f}|\psi_{k}^{i/f\pm}\rangle=\pm d_{k}^{i/f}|\psi_{k}^{i/f\pm}\rangle$,
with $d_{k}^{i/f}=\sqrt{[d_{x}^{i/f}(k)]^{2}+[d_{y}^{i/f}(k)]^{2}+[d_{z}^{i/f}(k)]^{2}}$.
Here we focus on the case in which ${\cal H}_{k}^{i}$ is Hermitian
with a gapped spectrum, and the initial state is a Slater determinant of lower band eigenstates $|\psi^{i-}_k\rangle$~\cite{supp4}. The translational symmetry allows us to treat the dynamics of each state $|\psi^{i-}_k\rangle$ separately.
The return amplitude is then given by $G^{-}(t)=\prod_{k}{\cal G}_{k}^{-}(t)$,
where ${\cal G}_{k}^{-}(t)=\cos\left(d_{k}^{f}t\right)-i\sin\left(d_{k}^{f}t\right)\langle\psi_{k}^{i-}|\frac{{\cal H}_{k}^{f}}{d_{k}^{f}}|\psi_{k}^{i-}\rangle$.
In the thermodynamic limit, the rate function of return probability reads

\begin{equation}
g^-(t)=-\frac{1}{2\pi}\int_{0}^{2\pi}dk\ln\left[|{\cal G}_{k}^{-}(t)|^{2}\right].\label{eq:RF}
\end{equation}
A DQPT is expected when a line of Lee-Yang-Fisher (LYF) zeros~\cite{LYF1,LYF2,LYF3}, defined as
\begin{equation}
z_{n,k}^{-}=i\frac{\pi}{d_{k}^{f}}\left(n+\frac{1}{2}\right)+\frac{1}{d_{k}^{f}}{\rm artanh}\langle\psi_{k}^{i-}|\frac{{\cal H}_{k}^{f}}{d_{k}^{f}}|\psi_{k}^{i-}\rangle\quad n\in\mathbb{Z},\label{eq:LYF}
\end{equation}
crosses the imaginary time axis at a critical momentum $k_{c}$, yielding
a critical time $t_{c,n}^{-}=-iz_{n,k_{c}}^{-}$ at which ${\cal G}_{k_{c}}^{-}(t_{c})=0$.
These conclusions hold for both Hermitian and non-Hermitian systems. 

To capture topological signatures of DQPTs in non-Hermitian systems,
we study the time dependence of the winding number
\begin{equation}
\nu_{{\rm D}}(t)=\frac{1}{2\pi}\int_{0}^{2\pi}dk\left[\partial_{k}\phi_{k}^{{\rm G}}(t)\right],\label{eq:WN}
\end{equation}
where the geometric phase of the return amplitude is given by $\phi_{k}^{{\rm G}}(t)\equiv\phi_{k}(t)-\phi_{k}^{{\rm dyn}}(t)$,
with the total phase $\phi_{k}(t)=-i\ln\left[\frac{{\cal G}_{k}^{-}(t)}{|{\cal G}_{k}^{-}(t)|}\right]$
and the dynamical phase
\begin{alignat}{1}
\phi_{k}^{{\rm dyn}}(t)= & -\int_{0}^{t}ds\frac{\langle\psi_{k}^{i-}(s)|{\cal H}_{k}^{f}|\psi_{k}^{i-}(s)\rangle}{\langle\psi_{k}^{i-}(s)|\psi_{k}^{i-}(s)\rangle}\nonumber \\
+ & \frac{i}{2}\ln\left[\frac{\langle\psi_{k}^{i-}(t)|\psi_{k}^{i-}(t)\rangle}{\langle\psi_{k}^{i-}(0)|\psi_{k}^{i-}(0)\rangle}\right].\label{eq:DYMP}
\end{alignat}
It can be shown that the geometric phase thus defined is {\it real} and
{\it gauge invariant}~\cite{supp2}.
These are the critical properties for us to introduce the real-valued, quantized
topological winding number $\nu_{\rm D}(t)$ for non-Hermitian systems.
Furthermore, the winding number of
$\phi_{k}^{{\rm G}}(t)$ in $k$-space will experience a quantized jump at
every critical time $t_{c}$ when the system passes through a DQPT point~\cite{WNJump}. As
will be shown, the value of $\nu_{{\rm D}}(t)$ will change monotonically
in time if the initial state is topologically trivial and if the quench
crosses an isolated EP.

\emph{Model.}--
To demonstrate our theory explicitly, we study a non-Hermitian lattice
model introduced in Ref.~\cite{NHTop13}, which may be realized in optical
resonators. In momentum representation, the model Hamiltonian is $H=\sum_{k}{\cal H}_{k}|k\rangle\langle k|$,
where
\begin{equation}
{\cal H}_{k}=h_{x}(k)\sigma_{x}+\left[h_{z}(k)+i\frac{\gamma}{2}\right]\sigma_{z},\label{eq:Hk}
\end{equation}
with $h_{x}(k)=\mu+r\cos k$ and $h_{z}(k)=r\sin k$. When the non-Hermitian
hopping amplitude $\gamma$ satisfies $\mu-r<\frac{\gamma}{2}<\mu+r$
or $\mu-r<-\frac{\gamma}{2}<\mu+r$, the trajectory of vector ${\bf h}(k)=[h_{x}(k),h_{z}(k)]$
versus quasimomentum $k$ will encircle one of the two EPs at ${\bf h}(k)=\left(\pm\frac{\gamma}{2},0\right)$.
In this case, the system is topologically nontrivial
and characterized by a winding number defined in a $4\pi$ Brillouin zone~\cite{NHTop13}.

We are going to study DQPTs following non-Hermitian quenches
in this lattice model. Specifically, we let ${\cal H}_{k}^{f}={\cal H}_{k}^{i}+i\frac{\gamma}{2}\sigma_{z}$,
where ${\cal H}_{k}^{i}=h_{x}(k)\sigma_{x}+h_{z}(k)\sigma_{z}$ is Hermitian.
The initial state $|\psi_{k}^{i}\rangle$ at each $k$ is chosen
to be the lower eigenstate of the prequench Hamiltonian ${\cal H}_{k}^{i}$
with energy $-d_{k}^{i}=-\sqrt{h_{x}^{2}(k)+h_{z}^{2}(k)}$. Dynamics
following a quench in which the vector ${\bf h}(k)$
encircles zero and one EP for both topologically trivial
and nontrivial initial states will be studied below~\cite{supp345}.
\begin{figure}
\includegraphics[trim=3.6cm 2.0cm 0.3cm 11.5cm, clip=true, height=!,width=1\columnwidth]{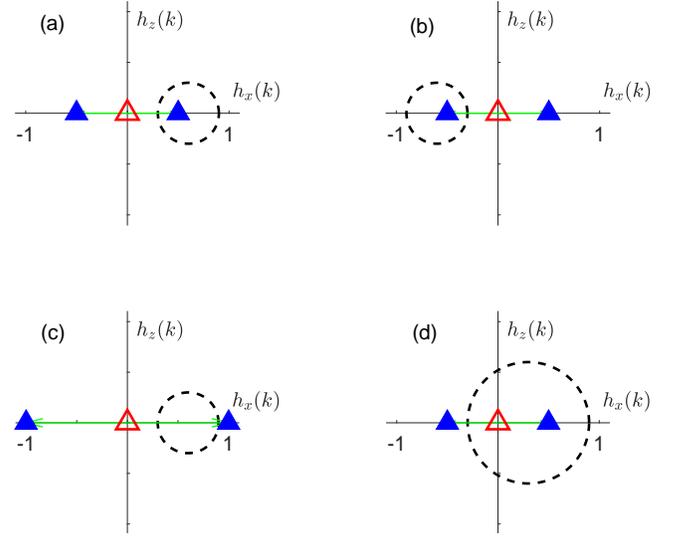}
\caption{(color online)
A sketch of the four quench cases. The hollow trangle denotes the zero of $h_x$-$h_z$ plane. The two solid triangles represent EPs of the Hamiltonian ${\cal H}^f_k$. The dashed circle represents the vector ${\bf h}(k)=[h_x(k),h_z(k)]$.
(a) Case 1: the initial state is topologically trivial, ${\bf h}(k)$ encircles the EP $\left(\frac{\gamma}{2},0\right)$.
(b) Case 2: the initial state is topologically trivial, ${\bf h}(k)$ encircles the EP $\left(-\frac{\gamma}{2},0\right)$.
(c) Case 3: the initial state is topologically trivial, ${\bf h}(k)$ does not encircle any EPs.
(d) Case 4: the initial state is topologically nontrivial, ${\bf h}(k)$ encircles the EP $\left(\frac{\gamma}{2},0\right)$.
  }
  \label{Protocol-1}
\end{figure}

\emph{Results}.--
Fig.~\ref{Protocol-1} gives a schematic of quench cases studied in this section.
In cases $1$, $2$ and $3$, the trajectory of ${\bf h}(k)$ does not encircle
the zero of $h_x$-$h_z$ plane, implying that the initial state is
topologically trivial~\cite{WNHermi}. In case $4$, the trajectory of ${\bf h}(k)$
encircles the zero of $h_x$-$h_z$ plane, and the initial state is
thus topologically nontrivial, characterized by a quantized winding
number. We now study the line of LYF zeros $z_{n,k}^{-}$, rate function
of return probability $g^-(t)$, geometric phase $\phi_{k}^{{\rm G}}(t)$
and winding number $\nu_{{\rm D}}(t)$ for nonunitary dynamics following
the four quench cases. These quantities provide with us key information
of DQPTs in the non-Hermitian lattice. 
\begin{figure}
	\includegraphics[trim=2.0cm 0.65cm 0.3cm 11.2cm, clip=true, height=!,width=1\columnwidth]{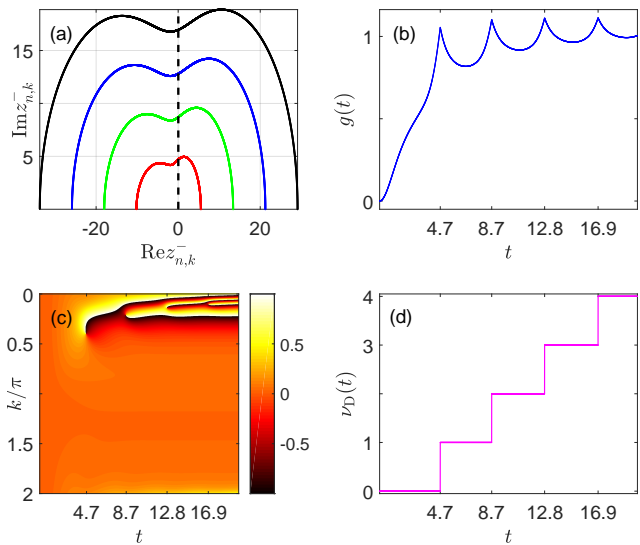}
	\caption{(color online) DQPTs for non-Hermitian quench case $1$. System parameters are $\mu=0.6$, $r=0.3$ for both pre- and postquench Hamiltonians, and $\gamma=1$.
		(a) Real and imaginary parts of lines of LYF zeros $z^-_{n,k}$ with $n=1,2,3,4$.
		(b) Rate function of the return probability~\cite{RGRate}. The first four DQPTs happen at $t\approx4.7,8.7,12.8$ and $16.9$.
		(c) Geometric phase $\phi^{\rm G}_k(t)$ versus quasimomentum $k$ and time $t$.
		(d) Time evolution of topological winding number $\nu_{\rm D}(t)$.
	}
	\label{Case-1}
\end{figure}

\begin{figure}
	\includegraphics[trim=2.0cm 0.65cm 0.3cm 11.2cm, clip=true, height=!,width=1\columnwidth]{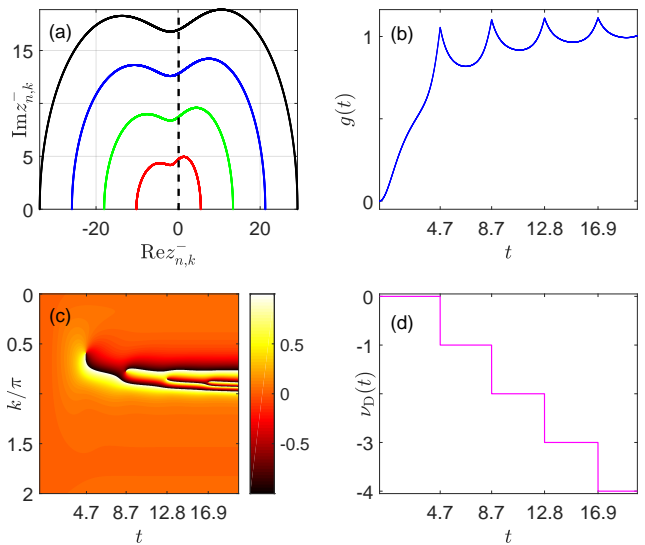}
	\caption{(color online) DQPTs for non-Hermitian quench case $2$. System parameters are $\mu=-0.6$, $r=0.3$ for both pre- and postquench Hamiltonians, and $\gamma=1$.
		(a) Real and imaginary parts of lines of LYF zeros $z^-_{n,k}$ with $n=1,2,3,4$.
		(b) Rate function of the return probability~\cite{RGRate}. The first four DQPTs happen at $t\approx4.7,8.7,12.8$ and $16.9$.
		(c) Geometric phase $\phi^{\rm G}_k(t)$ versus quasimomentum $k$ and time $t$.
		(d) Time evolution of topological winding number $\nu_{\rm D}(t)$.
	}
	\label{Case-2}
\end{figure}

Results of $z_{n,k}^{-}$, $g^{-}(t)$, $\phi_{k}^{{\rm G}}(t)$ and
$\nu_{{\rm D}}(t)$ for quench cases $1$ and $2$ are shown in Figs.~\ref{Case-1}
and \ref{Case-2}, respectively. In Fig.~\ref{Case-1}(a), we see that each line of LYF
zeros $z_{n,k}^{-}$ for $n=1,...,4$ has a unique crossing on the
imaginary time axis {[}{\it i.e.}, ${\rm Re}(z_{n,k}^{-})=0${]}, yielding
a positive critical time of DQPT at $t_{c,n}^{-}=-iz_{n,k_{c}}^{-}$ with
a unique critical momentum $k_{c}$. This is true for all $n>0$,
while there is no positive critical times for $n\leq0$. The same
pattern of LYF zeros is also observed for quench case $2$ as shown
in Fig.~\ref{Case-2}(a). In Figs.~\ref{Case-1}(b) and ~\ref{Case-2}(b), the rate functions behave non-analytically
at critical times $t_{c,n}^{-}$ found in Figs.~\ref{Case-1}(a) and \ref{Case-2}(a), which
is the defining feature of DQPTs. In Figs.~\ref{Case-1}(c) and \ref{Case-2}(c), the pattern
of geometric phase $\phi_{k}^{{\rm G}}(t)$ versus time $t$ and quasimomentum
$k$ is shown. We see that the winding number of $\phi_{k}^{{\rm G}}(t)$
in $k$-space increases every time when the evolution passes through
a critical time. This pattern is accurately accounted by the winding
number defined in Eq. (\ref{eq:WN}) for both quench cases,
as shown in Figs.~\ref{Case-1}(d) and \ref{Case-2}(d). Two notable features deserve to be mentioned.
\emph{First}, the value of $\nu_{{\rm D}}(t)$ is always quantized
and changes monotonically in time if the initial state is topologically
trivial and the postquench Hamiltonian vector ${\bf h}(k)$
encircles one of the EP. This provides with us a clear window to look
into the topological signature of an isolated EP in system's dynamics.
Following the classification scheme in unitary evolution, we refer
to DQPTs with quantized and monotonically changing winding numbers
as topologically nontrivial~\cite{HeylPRB2016_1}. \emph{Second}, the quench across the
EP in case $1$ induces a positive and monotonic increasing $\nu_{{\rm D}}(t)$
in time, while the quench across the EP in case $2$ induces a negative
and monotonic decreasing $\nu_{{\rm D}}(t)$. This difference originates
from the fact that the trajectory of vector ${\bf h}(k)$
in cases $1$ and $2$ crosses the branch cut connecting the two EPs
along opposite directions. We therefore find a way to distinguish
the two EPs from their topological response to a simple quench in
non-Hermitian systems, complementing other existing approaches focusing on
quasi-adiabatic evolution~\cite{NHDym3,NHDym4}. 

\begin{figure}
	\includegraphics[trim=2.0cm 8.8cm 0.3cm 11.2cm, clip=true, height=!,width=1\columnwidth]{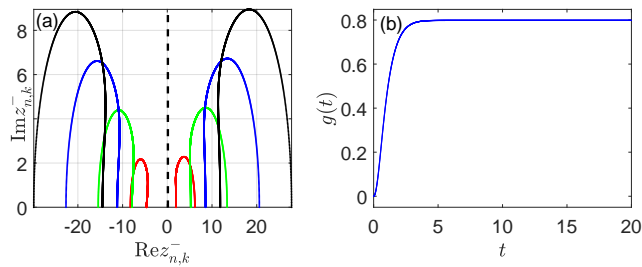}
	\caption{(color online) DQPTs for non-Hermitian quench case $3$. System parameters are $\mu=0.6$, $r=0.3$ for both pre- and postquench Hamiltonians, and $\gamma=2$.
		(a) Real and imaginary parts of lines of LYF zeros $z^-_{n,k}$ with $n=1,2,3,4$.
		(b) Rate function of the return probability~\cite{RGRate}.
	}
	\label{Case-3}
\end{figure}

In Fig.~\ref{Case-3}, we show results of $z_{n,k}^{-}$ and $g^{-}(t)$ for quench case $3$.
In this case, no EPs are encircled by the vector ${\bf h}(k)$, and
lines of LYF zeros have no crossings on the imaginary time axis as shown
in Fig.~\ref{Case-3}(a). This indicates that there is no DQPTs in this case and
the winding number remains zero in the nonunitary evolution, as also
supported by results reported in Fig.~\ref{Case-3}(b) and calculations of $\nu_{\rm D}(t)$.
More generally, for a vector ${\bf h}(k)$ encircling no EPs,
each line of LYF zeros could form a closed trajectory, crossing over
the imaginary time axis an even number of times~\cite{supp3}. This results in DQPTs
with a winding number oscillating in time, and thus can also be regarded
as topologically trivial~\cite{HeylPRB2016_1}.

\begin{figure}
	\includegraphics[trim=1.8cm 0.65cm 0.3cm 11.2cm, clip=true, height=!,width=1\columnwidth]{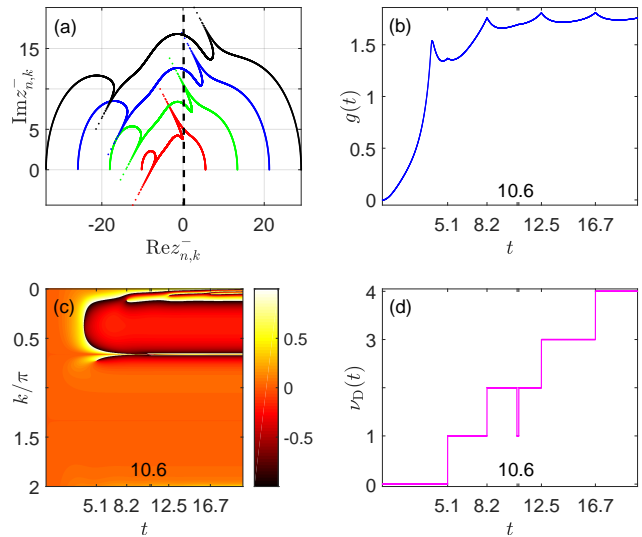}
	\caption{(color online) DQPTs for non-Hermitian quench case $4$. System parameters are $\mu=0.3$, $r=0.6$ for both pre- and postquench Hamiltonians, and $\gamma=1$.
		(a) Real and imaginary parts of lines of LYF zeros $z^-_{n,k}$ with $n=1,2,3,4$.
		(b) Rate function of the return probability~\cite{RGRate}.  The first four topological DQPTs happen at $t\approx5.1,8.2,12.5$ and $16.7$. A pair of accidental DQPTs happen at $t\approx10.6$.
		(c) Geometric phase $\phi^{\rm G}_k(t)$ versus quasimomentum $k$ and time $t$.
		(d) Time evolution of topological winding number $\nu_{\rm D}(t)$.
	}
	\label{Case-4}
\end{figure}

Results of $z_{n,k}^{-}$, $g^{-}(t)$, $\phi_{k}^{{\rm G}}(t)$ and
$\nu_{{\rm D}}(t)$ for quench case $4$ with a topologically nontrivial
initial state are presented in Fig.~\ref{Case-4}. Different from the first three cases of
trivial initial states, each line of LYF zeros now features two types
of crossing points on the imaginary time axis. One type of them appears
once for each line of LYF zeros regularly,
while the other type contains a pair of critical times separated by
a vanishingly small time window.
DQPTs corresponding to the later type of critical times may then be
regarded as originating from the memory of an evolving state about its
initial topology. Notably, this memory decays fast and becomes quickly
indistinguishable with the progress of time, leaving only signatures
of topologically nontrivial DQPTs following the quench across an EP.
These arguments are further supported by the time evolution of winding
number shown in Fig.~\ref{Case-4}(d), where a pair of accidental DQPTs caused
by the initial state topology appear at $t\approx10.6$. The insensitivity of
DQPTs in nonunitary evolution to the initial state topology has two
implications. \emph{First}, by choosing simple initial states, DQPTs due to
quenching across an EP can be demonstrated analytically in a transparent manner~\cite{supp5}.
\emph{Furthermore}, the non-Hermitian quench
may provide with us a useful strategy to generate topologically nontrivial
dynamical phases of matter from easy-to-prepare initial states.

In a photonic setup, the non-Hermitian quench may be engineered by introducing gain
or loss uniformly to half of the system along the propagation direction of the wave.
Then the signature of DQPTs may be identified by measuring and collecting phase factors
of each $k$-mode in space. The quantum walk setup reported in Ref.~\cite{NHTop5} could 
also be a candidate to experimentally study DQPTs in non-Hermitian systems.

\emph{Conclusion and discussion}.--
In summary, we have discovered DQPTs in non-Hermitian lattice systems by quenching
across EPs. The return probability vanishes when the nonunitary evolution
of the initial state reaches a critical time, where the topological
winding number displays a quantized jump. Two types of DQPTs were observed
and distinguished by the parity of critical times on each line of
LYF zeros. Topologically nontrivial DQPTs happen when the postquench
vector ${\bf h}(k)$ encircles an isolated EP. These transitions are characterized
by an odd number of critical times along each line of the LYF zeros,
and accompanied by a monotonically changing topological winding
number in time. Initial states with nontrivial topology can also leave
transient signatures in postquench dynamics, resulting in topologically
trivial accidental DQPTs. Whether the qualitative content of these findings hold
for more general dynamical protocols like linear ramp and periodic driving, and in
higher physical dimensions deserves further studies.

The concept of DQPT is initially introduced as an organizing principle to
study dynamics in closed nonequilibrium many-body systems. Discoveries made in this work could
serve as a starting point to generlize this concept to open quantum systems, whose dynamical
evolutions are usually nonunitary. The geometric phase and winding number introduced here, together
with their generalizations to density matrix formalism~\cite{Mixed1,Mixed2,supp5,SedlmayrOQS2017}, may also be useful tools to decode
geometric and topological order in the nonequilibrium dynamics of many-body open quantum systems.

In some studies concerning full counting statistics phase transitions,
quenching protocols with initial states prepared as right eigenvectors
of non-Hermitian spin chains are considered~\cite{HickeyPRB2014-1D}. The postquench Hamiltonian
there is Hermitian and the dynamics following the quench is unitary,
which is opposite to the situation considered in this paper. Also,
DQPTs observed there are mainly related to the
degeneracy point of a Hermitian system. In another study~\cite{NHTop1}, the time evolution of fidelity in a non-Hermitian
topological model is considered. There the focus is on coherence protection in finite-size systems.
Interestingly, zeros in the fidelity as a
function of certain time-dependent system parameters are also observed. The
relationship between these zeros and DQPTs may deserve
further explorations.

\vspace{1cm}

\emph{Acknowledgements.}-- L.Z. would like to thank Dajian Zhang and Linhu Li for useful discussions. Q.W. is supported by Singapore Ministry of Education Academic Research Fund Tier I (WBS No.~R-144-000-352-112). J.G. is supported by Singapore Ministry of Education Academic Research Fund Tier I (WBS No.~R-144-000-353-112) and by the Singapore NRF grant No.~NRF-NRFI2017-04 (WBS No.~R-144-000-378-281).


\begin{thebibliography}{99}
\bibitem{RGRate}
Here to show the nonanalytic behavior of $g^-(t)$ around DQPT points more clearly, we computed a renormalized rate function $g(t)\equiv-\frac{1}{2\pi}\int^{2\pi}_0 dk \ln[|{\cal G}^{-}_{k}(t)|^2/|e^{-i{\cal H}^f_k t}|\psi^{i-}_k\rangle|^2]$. This renormalized definition does not affect the description of qualitatively behaviors of the system around the DQPT point.
\end{thebibliography}
\end{document}